**The selfish ribosome**

Mart Krupovic[1]*, Eugene V. Koonin[2]*

Institut Pasteur, Université Paris Cité, Cell Biology and Virology of Archaea Unit, Paris, France.
Division of Intramural Research, National Library of Medicine, National Institutes of Health, Bethesda, MA, USA.

* - correspondence to:
Mart Krupovic, E-mail: mart.krupovic@pasteur.fr
Eugene V. Koonin, E-mail: koonin@ncbi.nlm.nih.gov

**Abstract**

The ribosome is responsible for protein synthesis in all cells, and is the cell's largest energy consumer. We propose that the ribosome originated as a mutualistic symbiont of an RNA-dependent RNA polymerase ribozyme, supplying peptides that enhanced replication. As life transitioned from the RNA to the RNA-protein world, autonomous replicators became irreversibly addicted to the ribosome for producing replication proteins. Subsequent evolution is construed as a ribosomal takeover, whereby the ribosome evolved to consume most of the cell's resources, while other cellular componentry ensured the propagation of the ribosome. Under this perspective, the ribosome is the ultimate biological selfish element.

**Introduction**

The origin of life is the most important unanswered question in biology. At the center of this problem is the ribosome, the main production facility of all cells that, together with other components of the translation system, is responsible for the synthesis of all proteins (Figure 1). As such, the ribosome is a quintessential, universal component of all cellular life forms [1, 2]. The ribosome is a massive and highly complex molecular machine that consists of two ribonucleoprotein (RNP) subunits, large (LSU) and small (SSU), which translocate the translated mRNA in a series of elaborate movements. Each subunit consists of a large RNA molecule (the LSU also contains a small RNA) complexed with multiple ribosomal (r) proteins, about 50, 65 and 80 r-proteins altogether in bacterial, archaeal and eukaryotic ribosomes, respectively. Among the about 100 genes that are universally conserved across cellular life, at least 95 encode RNA and protein components of the translation system (including several enzymes responsible for universal base modifications in rRNAs and tRNAs) [3]. Thus, the vertical inheritance of the translation system is readily traceable to the last universal cellular ancestor (LUCA), and accordingly, the sequences of rRNAs, r-proteins and other proteins involved in translation are used to build the universal phylogenetic tree (Tree of Life) and classify organisms into one of the three domains of life and lower taxa [4, 5]. The main motor of the ribosome, the peptidyltransferase center (PTC) in the LSU that catalyzes the formation of peptide bonds, is a *bona fide* ribozyme [6].

The universality of the ribosome and the ribozyme catalysis at its heart strongly suggest that the ribosome is a relic of the primordial RNA world [7, 8], arguably, the most prominent remnant of it surviving in modern cells. Thus, the ribosome has become the key study object for understanding the origin and early evolution of life [2, 9-11]. Detailed comparison of the ribosome structures prompted scenarios for the emergence of the ribosome and other components of the translation machinery from primordial RNA molecules [9, 12, 13]. The PTC, the most ancient component of the ribosome, apparently evolved through duplication of an RNA molecule capable of binding the CCA-3'-end of molecules that would later become tRNA [12], yielding the catalytically-competent proto-ribosome [14]. Subsequent evolution proceeded through gradual structural and functional complexification of the proto-ribosome [12, 13]. According to the "accretion model", LSU and SSU evolved progressively by accretion of RNA segments, with the SSU and proto-mRNAs (a population of single-stranded RNA molecules) initially functioning as cofactors that positioned and stabilized the interaction between LSU and tRNA [9]. The evolution of the ribosome appears to have been deciphered at considerable resolution, a truly remarkable feat given that this molecular machine is at least 4 billion years old [1, 11-13]. However, the biological or chemical environment, in which the ribosome evolved, and the evolutionary factors that made it the quintessential component of living organisms remain enigmatic.

The origin of life debate typically takes place within the framework of either the "metabolism first" or the "replication first" scenarios [15-17]. Under the "metabolism first" scenario, life evolved from increasingly complex autocatalytic chemical cycles, with genetic elements emerging at a later stage, whereas the "replication first" scenario holds that life started with the emergence and proliferation of genetic elements capable of self-replication (autonomous replicators) or replicating with the help of other elements (non-autonomous replicators) [18-20]. By contrast, emergence of the ribosome has not been central to the mainstream origin of life scenarios (even if it figures prominently in the discussion of the stages of evolution at the exit from the RNA world), with evolution of translation being considered a relatively late invention, postdating the emergence of replicators [18]. Indeed, the translation system with the ribosome at its heart can be plausibly perceived as a facilitator of replication and metabolism rather than the key actor of genetic inheritance and/or energy transformation in the cell. Here we present a different, ribosome-centric perspective on the evolution of life under which the ribosome is not merely the main workhorse of the cell, but rather the principal orchestrator of cellular activities, and the primary driver and beneficiary of life's evolution.

**The ribosome as the primary beneficiary of cellular processes**
Even the simplest modern cells are enormously complex, integrating structural components (cell envelope, membranes and cytoskeleton) with diverse systems involved in energy and chemical transformations (metabolism), and information processing systems (genome replication, transcription, translation). Although the ribosome is but one of many molecular machines in the cell, it is by far the most abundant one, quantitatively dwarfing all the others taken together. There are 20,000-27,000 ribosomes in each actively growing *Escherichia coli* cell [21, 22], whereas mammals contain up to 10 million ribosomes per cell [23]. Ribosomes occupy a substantial fraction of any cell's volume and are among the main contributors to molecular crowding, a key feature of the intracellular milieu [24]. Furthermore, the bacterial cell size limit is thought to be dictated by the "ribosome catastrophe", whereby the cell will require more ribosomes than can fit in its volume in order for the biosynthesis processes to keep up with the increasing growth rate [25]. In prokaryotes, that is, bacteria and archaea, rRNA represents ~95-98% of the total cellular RNA [26, 27], and the r-proteins collectively are the most abundant cellular proteins along with translation factors [28, 29]. Consequently, continuous production of ribosomes consumes a far greater share of the cell's energetic budget than any other process [28]. Specifically, translation has been estimated to account for at least 50% of the energy expenditure in a fast growing bacterial cell [30]. As pithily pointed out by Bowman and colleagues, "the function of a ribosome is to build more ribosomes" [2]. Thus, growth of a cellular population is, to a large extent, equivalent to the propagation of the ribosomes. Indeed, classic early studies by Jacques Monod's and Ole Maaloe's laboratories in the mid-20th century showed that the cellular concentration of ribosomes increases linearly with the growth rate [23]. More recent calculations demonstrated that ribosomal abundance and, specifically, the production time of each ribosome (~7 min in bacteria), set the hard limit on the achievable growth rate [31].

Traditionally, the heavy investment in the ribosome biogenesis and translation is considered from an organismal perspective, under the premise that high translation capacity begets faster cellular growth, increasing the cell's fitness. We ask, however, if an alternative perspective, whereby the ribosomal dominance in the cell reflects the selfish nature of the ribosome itself, could be equally or even more plausible? Can it be that cellular life from the early stages of evolution was hardwired to promote proliferation of the ribosomes and the evolution itself was largely driven by the "selfish interests" of the ribosomes?

**Coalescence of life around the ribosome**

Recent studies directed at the reconstruction of the ancestral ribosome have demonstrated non-templated ribozyme-catalyzed peptidyltransferase activity by dimers of PTC-containing segments of the LSU rRNA as small as 64 nucleotides (nt), suggesting that such ribozymes recapitulate the proto-ribosome [14, 32]. It is tempting to imagine that at the dawn of life, the ribozyme now located in the heart of the ribosome, the PTC, was also an RNA replicase responsible for the replication of the cognate (proto)rRNA itself and perhaps other RNA molecules [10, 33, 34]. However, there is no compelling evidence that such a ribozyme with a dual capacity of self-replication and templated translation or RNA molecules, with two separate, functionally distinct ribozymes ever existed. Until recently, the experimentally designed or evolved RNA polymerase have been complex molecules of at least 200 nt, much larger than the proto-ribosome, while still falling short of the catalytic efficiency, fidelity and processivity required of an RNA replicase [35-37]. It therefore appeared highly unlikely that the proto-ribosome could possess the additional RNA polymerase activity, making it far more plausible that the two activities evolved independently, in distinct RNA molecules [38]. The recent report of a 45 nt ribozyme polymerase capable of synthesizing itself and its complementary strand with a 94% fidelity (albeit from trinucleotide rather than mononucleotide substrates) could be a game changer, restoring the plausibility of a PTC-replicase ribozyme [39].

As proposed in a recent "symbiotic" model for the origin of life, the earliest autonomous and non-autonomous replicators likely formed symbiotic communities that existed in protocells (primordial reproducers), with some replicators being mutualists promoting the protocell reproduction and others being parasites [40]. Within such communities, the symbiosis between the two entities, one contributing the replication (replicase) and the other, translation (proto-ribosome) capacities could be mutualistic from the onset, with the proto-ribosome producing peptides, and eventually proteins, that enhanced the activity of the replicase (Figure 2). An alternative scenario would involve the dual-activity PTC-replicase ribozyme, with subsequent subfunctionalization [41] yielding dedicated proto-ribosome and replicase which would then coevolve as symbionts. The proto-ribosome, that is, the PTC ribozyme, would produce non-specific peptides some of which nonetheless could enhance the activity and/or stability of ribozyme replicases and other ribozymes. Given the ready availability of abiogenic amino acids, it appears likely that the PTC ribozyme was the first to emerge and that the RNA world was actually an RNA-peptide world from its very onset [42-44].

Beyond doubt, the ancestral translation system was much simpler in composition and functionality, compared to modern ribosomes, and likely produced peptides in a non-templated manner, from abiotically synthesized amino acids. However, emergence of this primitive translation system would likely give rise to a positive feedback loop [45] that, owing to the beneficial effect of the produced peptides, accelerated its own evolution involving complexification of the ribosome, evolution of translation factors [46, 47] and aminoacyl tRNA synthetases [48], and expansion of the genetic code, eventually establishing the modern-type, efficient, templated translation. Importantly, reconstruction of the evolution of common, ancient protein domains, such as the ubiquitous nucleotide-binding Rossmann fold, indicates that aminoacyl-tRNA synthetases, the key components of the modern translation system that define the specificity of amino acid incorporation into proteins, occurred relatively late in the evolutionary history of this domain [49, 50]. The emergence of the complex ribosome and the templated translation mark the transition from the RNA to the RNA-protein world [51] in which the replication of all replicators was catalyzed by proteins rather than by ribozymes. Following the switch in the replication strategy from ribozymes to protein enzymes and through production of other, increasingly diversifying proteins, the ribosome would solidify its indispensability for the survival of all replicators and the protocells themselves, akin to an addiction module (Figure 2).

The above scenario does not necessitate tight coupling between a specific type of replication machinery and the proto-ribosome, so that replication of the proto-rRNA could be performed by different co-evolving replicators encoding RNA-dependent RNA polymerases (RdRPs). There are at least two non-homologous RdRPs that can be traced to the early stages of evolution, namely, those containing the core RNA Recognition Motif (RRM) fold and those with the core double-psi beta barrel (DPBB)-fold [52]. Notably, both RRM and DPBB domains are structurally simple and could evolve from even simpler structural elements. For instance, the DPBB domain has a pseudo-two-fold symmetry, suggesting that it evolved from a simpler homodimer through duplication and gene fusion [53], mirroring the evolution of PTC. This hypothesis has been experimentally confirmed by reconstituting the DPBB fold from a peptide of ~40 amino acids [54]. Strikingly, it has been also demonstrated that through simple engineering experiments the DPBB fold could be transformed into RIFT barrel and OB folds, both common in ribosomal proteins and other components of the translation machinery [55, 56]. These experiments suggest that a broad repertoire of proteins involved in both replication and translation could evolve from the same initial structural scaffold through duplication, diversification and subfunctionalization, a major trend in the evolution of protein families. The multiplicity of primordial replicases is echoed in the contemporary life forms by the use of three non-homologous replicative DNA polymerases, one based on the RRM fold, another one on the DPBB fold, and the third one on the Polβ fold, by bacteria, archaea, eukaryotes and viruses [57]. The elaboration of the replicases by recruitment of additional proteins transforming them into highly efficient replisomes ultimately enabled the replication of much larger DNA genomes, in which the elements encoding the (proto)ribosome, the replicase, metabolic enzymes and other proteins were integrated into a single chromosome, thereby stabilizing their linkage and coevolution. The subsequent evolution of the early organisms proceeded along the same trajectory, via the positive feedback loop linking the continuous increase of the ribosome's efficiency with the ongoing emergence and refinement of new molecular systems and regulatory circuits.

Subsequent evolution of the ribosome, while preserving the core structure, involved a vast increase in size and complexity that occurred at early stages of evolution so that by the time of the LUCA, about 4 billion years ago, it has already reached the dimensions of the modern prokaryotic ribosomes (Figure 2) [1]. This process likely proceeded via hierarchical accretion of RNA elements and concomitant, stepwise addition of r-proteins [2, 12]. The growth and complexification of the ribosome is generally viewed as an adaptive process driven by selection for increased efficiency and elaborate regulation of translation. However, this massive accretion of structural elements transformed the ribosome not only into a highly efficient, tunable machine for protein synthesis but also into the main consumer of the resources of the cell. Indeed, the complexification of the proto-ribosome was inevitably accompanied by increasing energetic costs associated with its maintenance. Thus, the sustained proliferation of protocells containing both the proto-ribosome and the replicase required the evolution of an increasingly elaborate, robust and efficient metabolic network. In this framework, the evolution of the increasingly diverse metabolic pathways catalyzed by the (proto)ribosome-produced enzymes can be viewed as being driven primarily by the need to sustain the activity of the ribosome itself, a principle that continues to govern the functioning of modern cells. In this context, the ribosome is perhaps most pertinently viewed as a selfish entity subjugating the other functional systems of the cell.

**The ribosome and the multilevel selection paradigm**
The selfish ribosome concept naturally fits within the multilevel selection paradigm of the evolution of life including major and minor transitions in evolution [58-60]. Under this paradigm, selection operates at different levels simultaneously, and evolutionary transitions occur when a new, higher level of selection, or a new type of evolutionary units (individuality), emerges [61, 62]. The origin of cells, eukaryotes, multicellular life forms, and eusocial animals are all considered major transitions in evolution. In each transition, the units of selection at the lower, preceding level are combined into a new individual which becomes the principal unit of selection, whereas the lower level selection is largely but not completely suppressed. Frustration between competing selection processes at different levels of biological organization appears to be a key driver in the evolution of regulation, signal transduction and complexity in general [63].

The origin of multicellular organisms is the most obvious case in point. This transition occurred independently in many branches of the tree of life [64], and each time, most of the selection pressure was transferred from the level of individual cells to the level of the multicellular ensembles. Selection at the individual cell level does not cease, playing a role in development and particularly in immunity, but is tightly controlled through cell cycle checkpoints. Impairment of these checkpoints leads to an imbalance between selection at different levels and potentially to cancer [65, 66].

The origin of cells is the first and, in a sense, the most important major transition in evolution that is, obviously, much less thoroughly understood than the subsequent transitions. Nevertheless, there is no reasonable doubt that a key step in the evolution of modern-type cells was the integration of small genetic elements into large DNA genomes, with the agency of selection transferred from individual genes or small gene arrays to large ensembles of genes. Selection among small genetic elements did not stop as demonstrated by proliferation of numerous types of mobile elements (transposons, integrating conjugating elements and others) in the genomes of numerous, diverse life forms [67]. The selfish gene idea of Dawkins [68] and the even earlier similar ideas of Hamilton [69] preceded the concept of multilevel selection in its modern form but fit the framework perfectly. Indeed, genes can be thought of emerging as (semi)selfish replicators, with multilevel selection starting to operate at the early stage of protocells hosting symbiotic replicators [40] and the agency of selection shifting gradually from individual genes to growing gene collectives, culminating in large genomes comprising a single chromosome. Nevertheless, throughout the evolution of life, genes have retained selfish features as emphatically illustrated by horizontal spread of successful genes and 'selfish' operons [70-72] across the prokaryote world.

Within this paradigm, the ribosome and the translation system appears as the ultimate manifestation of selfishness combined with 'altruistic' behavior. Indeed, the translation system consumes the bulk of the cell's energy and chemical building blocks, exploiting the rest of the cellular functional systems while supplying them with essential proteins. Furthermore, genes and operons encoding rRNAs and tRNAs are amplified in most organisms [73, 74], demonstrating their evolutionary success that also promotes the success of the cells and organisms. Remarkably, rRNA operons are occasionally located on plasmids, *bona fide* selfish elements that promote rRNA proliferation and spread [75]. Under the "selfish ribosome" concept, the (proto)ribosome was the primary unit of selection from the onset of cellular life and remained such throughout cellular evolution. The selection pressure on other cellular systems, including replicators and metabolic networks, can be viewed as relaxed, facilitating their flux and exchange between cellular lineages across the tree of life. Hence the striking constancy of the translation system throughout the 4 billion years of the evolution of life, in stark contrast to the fluidity of the replication machinery and metabolism.

**The ribosome at the heart of biological conflicts**
The selfish ribosome concept offers a distinct perspective not only on the origin and evolution of cellular life forms, but also on the interaction between cells and selfish replicators, most notably, viruses. Although viruses with large genomes often encode various components of the translation machinery, including tRNA genes, translation factors, tRNA synthetases and occasionally even r-proteins [76-78], no virus capable of autonomous translation has been discovered thus far. Indeed, cells and viruses have been denoted as ribosome-encoding organisms and capsid-encoding or virion-encoding organisms, respectively [79, 80]. Thus, selfish replicators, which eventually gave rise to *bona fide* viruses upon acquiring the ability to form virus particles and move between the cells extracellularly [81], just like autonomous replicators, became addicted to the ribosome, establishing the inviolable, billion-years old dependence of viruses on their hosts. Even RNA-based, non-coding replicators, such as viroids, which are often considered to be relics of the RNA world, depend on proteins produced by the ribosome for replication [82]. Of note, virus genome replication and assembly of infectious virions can be achieved for many RNA and DNA viruses in the absence of intact cells, provided that all components of the translation system and the necessary building blocks are supplied [83, 84]. The cell-free virus production emphasizes the basic dependence of viruses on the translation machinery but not necessarily on the overall cellular integrity.

Conceivably, viruses do not encode their own ribosomes due to restrictive energy and temporal costs. Given the sheer number of rRNA and protein molecules constituting the ribosome, the complexity of their assembly involving multiple chaperones and modification enzymes, and the number of ribosomes required for efficient translation, the energetic expenditure of supplanting the host translation machinery with a viral one would be prohibitive. Besides, production of a single bacterial ribosome takes ~7 min [31] and thus *de novo* production of viral ribosomes at the onset of infection would be temporally detrimental for efficient virus reproduction and spread in the population. Consequently, hijacking the host translation apparatus appears to be the only viable strategy for genetic parasites. Indeed, viruses have evolved many ways to subvert nearly every step in the host translation process [85-87]. Reciprocally, cells have devised various antiviral strategies at the level of translation, restricting viral access to the ribosomes [85]. Notably, numerous bacterial toxin-antitoxin systems, which function in antiviral defense, target all steps of translation via diverse mechanisms [88-90].

The translation machinery is also at the heart of inter-cellular conflicts, whereby the ribosome is the prime target of numerous antibiotics and antimicrobial peptides [91-94], whereas resistance evolves via various routes including diverse rRNA modifications [95, 96]. Thus, as befits a quintessential selfish entity, the ribosome is the fulcrum of biological conflicts at multiple levels. The study of these conflicts at the ribosomal interface provides new insights not only into molecular mechanisms of translation and virus-host interactions but also the co-evolutionary arms races between the perennial adversaries, the (selfish) ribosomes and the (even more) selfish replicators, as well as between competing microbes.

**Conclusion**
Here we propose a perspective on the origin and evolution of life where the ribosome and the rest of the translation machinery take the center stage in cellular functionality and evolution, as a distinct type of selfish entity that consumes the bulk of the cell's resources, subjugating the rest of cellular systems while providing them with proteins essential for their functions. The coexistence of the translation, transcription and replication systems with the operational componentry of the cell is strictly mutualistic. It is therefore impossible to determine objectively whether the cell serves the ribosome or *vice versa*. However, the selfish ribosome perspective is attractive in that it connects the translation machinery that is universal and central to all modern cells with the primordial stage of evolution. At that stage, symbiosis between different types of replicators including the proto-ribosome and the primordial replicases could have driven the transition from protocells populated by small, (quasi)selfish replicators to the full-fledged, modern-type cells endowed with large DNA genomes consolidating all genes on a single or a few chromosomes. Under the multilevel selection paradigm, the selfish character of the ribosome was masked during the evolution of cellular life forms but still lurks in modern organisms.

**Author contributions:** M.K. and E.VK. jointly developed the concept and wrote the manuscript.

**Acknowledgements:** The authors thank Purificación López-García, Michael Lynch and Andrew Ellington for insightful discussions.

**Funding:** M.K. was supported by the European Union HORIZON-EIC-2024 project VirHoX. E.V.K was supported by the Intramural Research Program of the National Institutes of Health of the USA. The funders had no role in study design, data collection and analysis, decision to publish, or preparation of the manuscript.

**Competing interests:** The authors have declared that no competing interests exist.

**Figures**

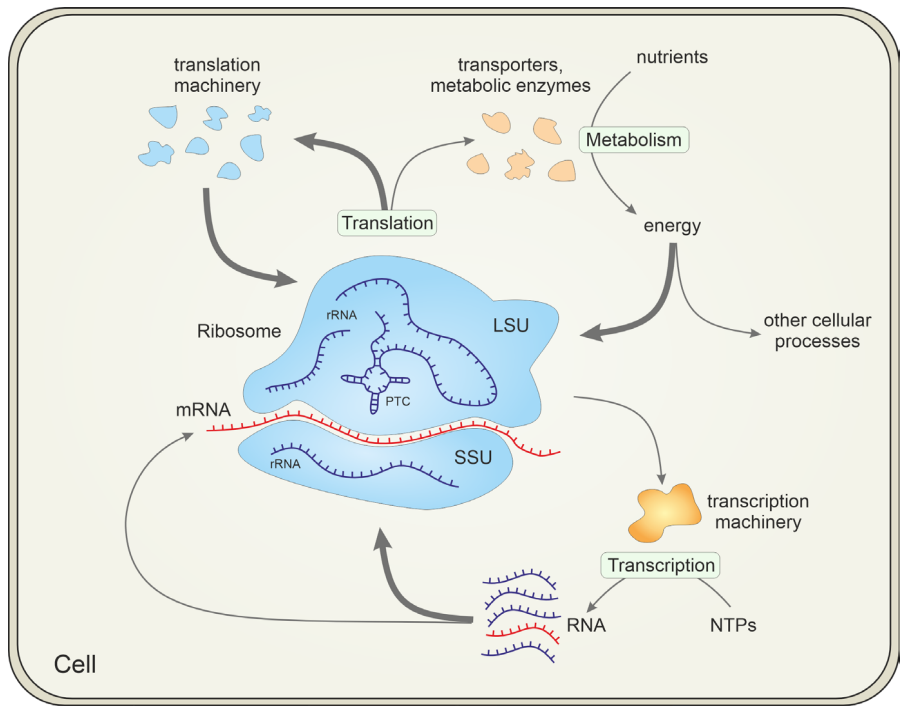

**Figure 1. The central place of the ribosome in the cell**. The figure shows that ribosomes consume the largest share (indicated with thick arrows) of energy, transcribed RNA and translated proteins. Abbreviations: rRNA, ribosomal RNA; mRNA, messenger RNA; SSU and LSU, small and large subunits of the ribosome, respectively; PTC, peptidyltransferase center; NTPs, nucleoside triphosphates.

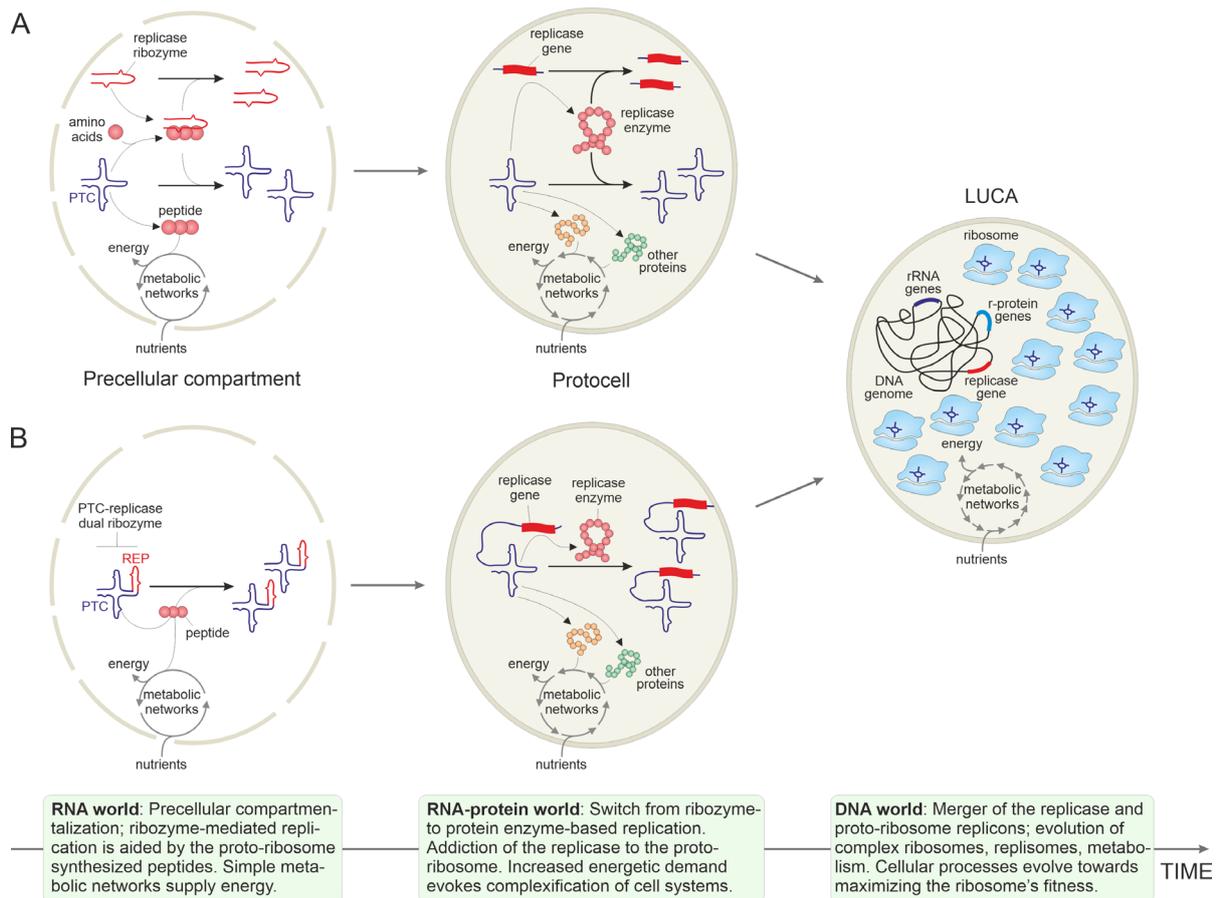

**Figure 2. A symbiotic scenario for co-evolution of the ribosome and the replicase.** Two variations of the scenario are depicted, whereby the peptidyltransferase center (PTC) and the replicase (REP) ribozymes are encoded by distinct (A) or the same (B) RNA molecule. The figure depicts three major stages of cellular evolution from the primordial RNA world (left) to the intermediate RNA-protein world (center) and the last universal cellular ancestor (LUCA, right). The major events occurring during each stage are briefly described in the corresponding information boxes at the bottom of the figure.